\documentclass[12pt]{iopart}

\usepackage{graphicx}
\usepackage{dcolumn}
\usepackage{bm}
\newcommand{\x}{{\bm x}}
\newcommand{\be}{\begin{equation}}
\newcommand{\ee}{\end{equation}}
\begin{document}


\title{Hydrokinetic simulations of nanoscopic precursor films in rough channels} 

\author{S. Chibbaro$^{1}$, L. Biferale$^{2}$, K. Binder$^{3}$, D. Dimitrov$^{4}$,F. Diotallevi$^{5}$, A. Milchev$^{3}$, S. Succi$^{5}$ \\
$^{1}$  Dept. of Mechanical Engineering, University of ``Tor Vergata'', via del politecnico 1 00133, Rome, Italy \\
$^{2}$ Dipartimento di Fisica e INFN, Universita' di Tor Vergata, Via della Ricerca Scientifica 1, 00133 Rome, Italy\\  
$^{3}$  Institut fur Physik, Johannes Gutenberg Universitat Mainz, Staudinger Weg 7, 55099 Mainz Germany \\
$^{4}$ University of Food Technologies, 26 Maritza
 blvd., 4000 Plovdiv, Bulgaria \\
$^{5}$ Istituto per le Applicazioni del Calcolo CNR  V. Policlinico 137, 00161 Roma, Italy }

\begin{abstract}  We report  on  simulations of  capillary filling  of
high-wetting fluids  in nano-channels  with and without  obstacles. We
use     atomistic     (molecular     dynamics)    and     hydrokinetic
(lattice-Boltzmann) approaches  which point out clear  evidence of the
formation of thin precursor films,  moving ahead of the main capillary
front.   The  dynamics of  the  precursor films  is  found  to obey  a
square-root  law as  the  main capillary  front,  $z^2(t) \propto  t$,
although with a larger prefactor, which we find to take the same value
for  the  different  geometries  (2D-3D) under  inspection.   The  two
methods  show  a  quantitative  agreement  which  indicates  that  the
formation  and propagation  of thin  precursors  can be  handled at  a
mesoscopic/hydrokinetic level. This can  be considered as a validation
of  the Lattice-Boltzmann  (LB) method  and opens  the  possibility of
using hydrokinetic  methods to  explore space-time scales  and complex
geometries  of direct  experimental relevance.   Then, LB  approach is
used to study the fluid behaviour in a nano-channel when the precursor
film encounters a square  obstacle.  A complete parametric analysis is
performed  which  suggests  that  thin-film  precursors  may  have  an
important   influence  on   the   efficiency  of   nanochannel-coating
strategies.
\end{abstract}

\maketitle

\section{Introduction}   Micro-   and   nano-hydrodynamic  flows   are
prominent  in many  applications  in material  science, chemistry  and
biology~\cite{DeG_84,DeG_book,Bhu_95,Zha_01,Whi_01,Str_02,Mas_02,Sup_03,Yan_04,Jos_06,Jua_06,Yak_07}. A
thorough  fundamental  understanding as  well  as  the development  of
corresponding  efficient   computational  tools  are   demanded.   The
formation of thin precursor films in capillary experiments with highly
wetting fluids (near-zero contact angle) has been reported by a number
of               experiments              and              theoretical
works~\cite{Bonn,Kav_03,Bico,Blake,Heslot}, mostly  in connection with
droplet spreading, and only  very recently~\cite{Chi_08b} for the case
of capillary filling.   In this latter case the  presence of precursor
films could help  to reduce the drag, and this  could have an enormous
economic impact as mechanical technology is miniaturized, microfluidic
devices  become  more  widely  used,  and  biomedical  analysis  moves
aggressively towards  lab on a chip technologies.   In this direction,
patterned channels~\cite{Kus}  and more specifically ultra-hydrophobic
surfaces have  been considered~\cite{Ou,Hyv}.  At  microscopic scales,
inertia  subsides   and  fluid  motion  is  mainly   governed  by  the
competition   between   dissipation,   surface-tension  and   external
pressure.   In  this  realm,   the  continuum  assumption  behind  the
macroscopic  description  of fluid  flow  goes  often under  question,
typical cases  in point  being slip-flow at  solid walls and  moving a
contact     line     of      liquid/gas     interface     on     solid
walls~\cite{DeG_84,Bonn}.  In order to keep a continuum description at
nanoscopic  scales  and  close  to the  boundaries,  the  hydrodynamic
equations are  usually enriched with  generalised boundary conditions,
designed in such a way as to collect the complex physics of fluid-wall
interactions into a few effective  parameters, such as the slip length
and the  contact angle~\cite{Troian,Red}.  A more  radical approach is
to  quit  the continuum  level  and  turn  directly to  the  atomistic
description   of    fluid   flows   as   a    collection   of   moving
molecules~\cite{Allen_Tildesley},   typically    interacting   via   a
classical   6-12   Lennard-Jones    potential.    This   approach   is
computationally demanding, thus preventing the attainment of space and
time  macroscopic scales  of  experimental interest.   In between  the
macroscopic  and microscopic  vision, a  mesoscopic approach  has been
lately  developed in  the  form  of minimal  lattice  versions of  the
Boltzmann        kinetic       equation~\cite{SS_92,wolf}.        This
mesoscopic/hydro-kinetic approach offers  a compromise between the two
methods,  i.e.   physical  realism  combined with  high  computational
efficiency.  By definition, such  a mesoscopic approach is best suited
to situations where molecular details, while sufficiently important to
require  substantial  amendments of  the  continuum assumption,  still
possess a sufficient degree of universality to allow general continuum
symmetries   to  survive,  a   situation  that   we  shall   dub  {\it
supra-molecular}  for   simplicity.   Lacking  a   rigorous  bottom-up
derivation,   the   validity  of   the   hydro-kinetic  approach   for
supra-molecular physics must be assessed case-by-case, a program which
is    already    counting     a    number    of    recent    successes
\cite{SS_06,Sbr_06,Yeom,Harting}.  The aim  of this paper is two-fold.
First, we validate the  Lattice-Boltzmann (LB) hydro-kinetic method in
another  potentially 'supramolecular'  situation, i.e.   the formation
and propagation of precursor  films in capillary filling at nanoscopic
scales.   We  employ both  MD  and  hydrokinetic simulations,  finding
quantitative  agreement for  both  bulk quantities  and local  density
profiles, at all times during the capillary filling process.  Then, we
carry out  a complete LB study of  a nanochannel in the  presence of a
square  obstacle  for  a  large  range  of  value  for  the  different
parameters at  play.  These results  seem to suggest that,  by forming
and  propagating   ahead  of  the  main   capillary  front,  thin-film
precursors may manage to  hide the chemical/geometrical details of the
nanochannel  walls,   thereby  exerting  a  major   influence  on  the
efficiency  of nanochannel-coating strategies  \cite{Cottin,tas}.  The
paper is  organised as follows: in  the first two  sections the models
employed are presented with some details. In section IV, the numerical
results are discussed: first the study of a nanochannel for a complete
wetting is  presented and, then, the  LB analysis of  a nanochannel in
presence of  a square obstacle  is reported. Finally,  conclusions are
made.

\section{Lattice-Boltzmann   model}   In   this   work  we   use   the
multicomponent LB model proposed  by Shan and Chen \cite{SC_93}.  This
model  allows for  distribution functions  of an  arbitrary  number of
components, with different molecular mass:
\begin{equation}
\label{eq:lbe} f^k_i(\x+ {\bm c}_{i} \Delta t,t+\Delta t)-f^k_i(\x,t)=
-\frac{ \Delta t}{\tau_k} \left[f^k_i(\x,t) -f_i^{k(eq)}(\x,t)\right]
\end{equation}  where  $f^k_i(\bm{x},t)$  is the  kinetic  probability
density  function associated with  a mesoscopic  velocity $\bm{c}_{i}$
for the  $k$th fluid, $\tau_k$ is  a mean collision time  of the $k$th
component (with $\Delta t$ a time step), and $f^{k(eq)}_{i}(\x,t)$ the
corresponding equilibrium function.   The collision-time is related to
kinematic          viscosity          by          the          formula
$\nu_k=\frac{1}{3}(\tau_k-\frac{1}{2})$.     For   a   two-dimensional
9-speed LB model (D2Q9) $f^{k(eq)}_{i}(\x,t)$ takes the following form
\cite{wolf}:
\begin{eqnarray}
\label{eq:feq1}          f^{k(eq)}_{0}&=&\alpha_kn_k-\frac{2}{3}n_k{\bf
u}_k^{eq}\cdot{\bf                     u}_k^{eq}                    \\
f^{k(eq)}_{i}&=&\frac{(1-\alpha_k)n_k}{5}+\frac{1}{3}n_k{\bf
c}_i\cdot{\bf  u}_k^{eq}  \label{eq:feq2}  \\ &+&  \frac{1}{2}n_k({\bf
c}_i\cdot{\bf    u}_k^{eq})^2-\frac{1}{6}n_k{\bf    u}_k^{eq}\cdot{\bf
u}_k^{eq}     \;\;\;\textrm{for     i=1$\ldots$4}     \nonumber     \\
f^{k(eq)}_{i}&=&\frac{(1-\alpha_k)n_k}{20}+\frac{1}{12}n_k{\bf
c}_i\cdot{\bf  u}_k^{eq}  \label{eq:feq3}  \\ &+&  \frac{1}{8}n_k({\bf
c}_i\cdot{\bf    u}_k^{eq})^2-\frac{1}{24}n_k{\bf   u}_k^{eq}\cdot{\bf
u}_k^{eq} \;\;\;\textrm{for i=5$\ldots$8} \nonumber
\end{eqnarray}  In  the above  equations  ${\bf  c}_i$'s are  discrete
velocities, defined  as follows \be {\bf c}_i=  \left \{ \begin{array}
{l}                    0,                    i=0,                   \\
\left(cos\frac{(i-1)\pi}{2},sin\frac{(i-1)\pi}{2}\right),   i=1-4   \\
\sqrt{2}\left(cos[\frac{(i-5)\pi}{2}+\frac{\pi}{4}],sin[\frac{(i-5)\pi}{2}+\frac{\pi}{4}]\right),
i=5-8
\end{array} \right.  \ee
where $\alpha_k$ is a free parameter
related  to the  sound  speed  of the  $k$th  component, according  to
$(c_s^k)^2=\frac{3}{5}(1-\alpha_k)$;  $n_k=\sum_if^k_i$ is  the number
density  of  the $k$th  component.  The  mass  density is  defined  as
$\rho_k=m_kn_k$, and the fluid velocity of the $k$th fluid ${\bf u}_k$
is  defined through  $\rho_k{\bf u}_k=m_k\sum_i{\bf  c}_if_i^k$, where
$m_k$ is the  molecular mass of the $k$th  component.  The equilibrium
velocity ${\bf u}_k^{eq}$ is determined by the relation \be \rho_k{\bf
u}_k^{eq}=\rho_k {\bf U}  +\tau_k{\bf F}_k \ee where ${\bf  U}$ is the
common velocity of  the two components.  To conserve  momentum at each
collision in  the absence  of interaction (i.e.  in the case  of ${\bf
F}_k=0$)   ${\bf  U}$   has   to  satisfy   the   relation  \be   {\bf
U}=\left(\sum_i^s        \frac{\rho_k{\bf        u}_k}{\tau_k}\right)/
\left(\sum_i^s  \frac{\rho_k}{\tau_k}\right)\;.   \ee The  interaction
force between  particles is the sum  of a bulk and  a wall components.
The bulk force is given by \be
\label{forcing}  {\bf  F}_{1k}({\bf  x})=-\Psi_k(\x)\sum_{\x^{\prime}}
\sum_{\bar{k}=1}^sG_{k                           \bar{k}}\Psi_{\bar{k}}
(\x^{\prime})(\x^{\prime}-\x)  \ee where  $G_{k\bar{k}}$  is symmetric
and   $\Psi_k$  is   a  function   of  $n_k$.    In  our   model,  the
interaction-matrix    is   given    by    \be   G_{k    \bar{k}}=\left
\{  \begin{array}  {l}  g_{k  \bar{k}},  |\x^{\prime}-\x|=1,  \\  g_{k
\bar{k}}/4,          |\x^{\prime}-\x|=\sqrt{2},          \\         0,
\textrm{otherwise}. \end{array} \right.   \ee where $g_{k \bar{k}}$ is
the strength of the inter-particle potential between components $k$ and
$\bar{k}$. In  this study, the effective  number density $\Psi_k(n_k)$
is taken  simply as $\Psi_k(n_k)=n_k$.  Other choices would lead  to a
different equation of state (see below).

At the  fluid/solid interface,  the wall is  regarded as a  phase with
constant number density.  The  interaction force between the fluid and
wall is described as \be
\label{forcingw} {\bf F}_{2k}({\bf x})=-n_k(\x)\sum_{\x^{\prime}} g_{k
w}  \rho_{w} (\x^{\prime})(\x^{\prime}-\x) \ee  where $\rho_w$  is the
number density  of the wall  and $g_{kw}$ is the  interaction strength
between component $k$ and the wall. By adjusting $g_{kw}$ and $\rho_w$
, different  wettabilities can be obtained.  This  approach allows the
definition  of  a static  contact  angle  $\theta$,  by introducing  a
suitable value for the  wall density $\rho_w$ \cite{Kan_02}, which can
span the  range $\theta  \in [0^o:180^o]$.  
In that work~\cite{Kan_02}, for the first time to the best of our knowledge, this phenomenological definition of the contact angle was put forward and 
the multi-component lattice Boltzmann method was used to study the displacement of a two-dimensional immiscible
droplet subject to gravitational forces in a channel. In particular, the dynamic behavior of the droplet was analysed,
and the effects of the contact angle, Bond number (the ratio of gravitational to surface forces),
droplet size, and density and viscosity ratios of the droplet to the displacing fluid were investigated.
It  is worth  noting that,
with this method, it is not possible to know ``a priori'' the value of
the contact angle from  the phenomenological parameters.  Thus, an ``a
posteriori'' map of  the value of the static  contact angle versus the
value of the  interaction strength $g_w$ has to  be obtained.  To this
aim,  we have  carried out  several  simulations of  a static  droplet
attached to a wall for different values of $g_w$~\cite{Kan_02,ben_06}.
In particular, in our work, the  value of the static contact angle has
been  computed directly  as the  slope  of the  contours of  near-wall
density field, and independently through the Laplace's law, $\Delta P=
\frac{2 \gamma cos  \theta}{H}$, where $H$ is the  channel height.  The
value  so  obtained  is  computed  within  an  error  $\sim  2\%-3\%$.
Recently, a  different approach  has been proposed,  which is  able to
give an  ``a priori''  estimate of the  static contact angle  from the
phenomenological   parameter~\cite{Hua_07}.   Nevertheless,   we  have
preferred to retain our ``a posteriori'' method for its simplicity and
efficiency.

In  a  region  of pure  $k$th  component,  the  pressure is  given  by
$p_k=(c_s^k)^2m_kn_k$,  where $(c_s^k)^2=\frac{3}{5}(1-\alpha_k)$.  To
simulate a  multiple component fluid with different  densities, we let
$(c_s^k)^2m_k=c_0^2$,  where $c_0^2=1/3$.  Then,  the pressure  of the
whole             fluid             is            given             by
$p=c_0^2\sum_kn_k+\frac{3}{2}\sum_{k,\bar{k}}g_{k,\bar{k}}\Psi_k\Psi_{\bar{k}}$,
which represents a non-ideal gas law.

The Chapman-Enskog expansion \cite{wolf}  shows that the fluid mixture
follows the Navier-Stokes equations for a single fluid:
\begin{eqnarray}
\partial_{t}\rho  +  \nabla  \cdot   (\rho  {\bf  u})  &=&0,  \\  \rho
[ \partial_{t} {\bm  u} + ({\bm u} \cdot {\bm \nabla}  ){\bm u}] &=& -
{\bf \nabla} {P} +{\bm F} + {\bm \nabla} \cdot (\mu( {\bm \nabla} {\bm
u}+{\bm u}{\bm \nabla})) \nonumber
\label{eq:NS}
\end{eqnarray} where $\rho=\sum_k \rho_k$  is the total density of the
fluid  mixture, the  whole  fluid  velocity ${\bf  u}$  is defined  by
$\rho{\bf u}=  \sum_k \rho_k{\bf u}_k+\frac{1}{2}\sum_k{\bf  F}_k$ and
the  dynamic viscosity  is  given  by $\mu=  \rho  \nu= \sum_k\mu_k  =
\sum_k(\rho_k\nu_k)$.

To  the purpose of  analysing the  physics of  film precursors,  it is
important to notice that ,  in the limit where the hard-core repulsion
is negligible,  both the Shan-Chen pseudo-potential and  Van der Waals
interactions  predict a  non-ideal  equation of  state,  in which  the
leading  correction  to  the   ideal  pressure  is  $\propto  \rho^2$.
Therefore, both models obey the Maxwell area rule ~\cite{ben_06}.

\section{MD  model} 

In the  Molecular Dynamics  simulation we  use the
simplest  model, consisting  of a fluid of point-size particles
that interact via a  Lennard-Jones potential. 
Henceforth all lengths will be quoted in units
of  $\sigma$, the atom diameter.
The  snapshot in
Fig.~\ref{Fig:1}b, illustrates our  simulation geometry. We consider a
cylindrical nanotube  of radius $R=11$ and length  $L=80$, whereby the
capillary  walls are  represented by  densely packed  atoms  forming a
triangular lattice with lattice constant  $1.0$.
  The wall  atoms may  fluctuate around  their equilibrium
positions, subjected  to a finitely extensible non-linear elastic
(FENE) potential,
\begin{equation}\label{FENE} U_{FENE}=-15\epsilon_w R_0^2 \ln\left (1-
r^2/R_0^2\right ),\;R_0=1.5
\end{equation} Here $r$  is the distance between the  particle and the
virtual  point  which  represents  the  equilibrium  position  of  the
particle in  the wall structure, $\epsilon_w=1.0  k_BT$, $k_B$ denotes
the Boltzmann constant,  and $T$ is the temperature  of the system. In
addition, the wall atoms interact by a Lennard-Jones (LJ) potential,
\begin{equation}
\label{LJ}
U_{LJ}(r)=4\epsilon_{ww}
\left[(\sigma_{ww}/r)^{12}-(\sigma_{ww}/r)^6\right],
\end{equation}  
where $\epsilon_{ww}=1.0$ and  $\sigma_{ww}=0.8$. This
choice of  interactions guarantees no penetration  of liquid particles
through the wall while in the same time the mobility of the wall atoms
corresponds  to the  system  temperature. The particles  of the  liquid
interact  with each  another by  a LJ-potential  with  $\epsilon_{ll} =
1.40$ so that the resulting fluid attains a density of $\rho_l \approx
0.77$. The  liquid film is in  equilibrium with its vapor  both in the
tube  as   well  as  in  the   partially  empty  right   part  of  the
reservoir. The  interaction between fluid particles and  wall atoms is
also described  by a Lennard-Jones potential,  Eq.~(\ref{LJ}), of range
$\sigma_{wl}=1$ and strength $\epsilon_{wl}=1.4$.

Molecular Dynamics (MD) simulations  were performed using the standard
Velocity-Verlet  algorithm \cite{Allen_Tildesley} with  an integration
time step $\delta  t = 0.01 t_0$ where the MD  time unit (t.~u.) $t_0=
(\sigma  ^2m /  48 \epsilon_{LJ})^{1/2}=1/\sqrt{48}$  and the  mass of
solvent particles $m=1$.  Temperature was held constant at $T=1$ using
a   standard   dissipative    particle   dynamics   (DPD)   thermostat
\cite{Soddemann,DPD}  with a  friction constant  $\zeta =  0.5$  and a
step-function like weight function  with cutoff $r_c=1.5 \sigma$.  All
interactions are  cut off at  $r_{cut}=2.5\sigma$. The time  needed to
fill the capillary is of the order of several thousands MD time units.

The top of the capillary is closed by a hypothetical impenetrable wall
which prevents liquid atoms escaping from the tube.  At its bottom the
capillary is attached to a rectangular $40\times 40$ reservoir for the
liquid with  periodic boundaries perpendicular  to the tube  axis, see
Fig.~\ref{Fig:1}b.   Although  the liquid  particles  may move  freely
between  the reservoir  and the  capillary tube,  initially,  with the
capillary walls being taken distinctly lyophobic, these particles stay
in the reservoir as a thick  liquid film which sticks to the reservoir
lyophilic right wall.  At time $t=0$, set to be the onset of capillary
filling,  we  switch   the  lyophobic  wall-liquid  interactions  into
lyophilic  ones  and  the  fluid  enters the  tube.  Then  we  perform
measurements  of   the  structural  and  kinetic   properties  of  the
imbibition process  at equal intervals  of time.  The total  number of
liquid  particles is  $4\times  10^5$ while  the  number of  particles
forming the tube is $4800$.

\section{Numerical Results}

\subsection{Nanochannel capillary filling with complete wetting}

\begin{figure} \begin{center} \vspace{0.5cm} (a)
\includegraphics[scale=0.35]{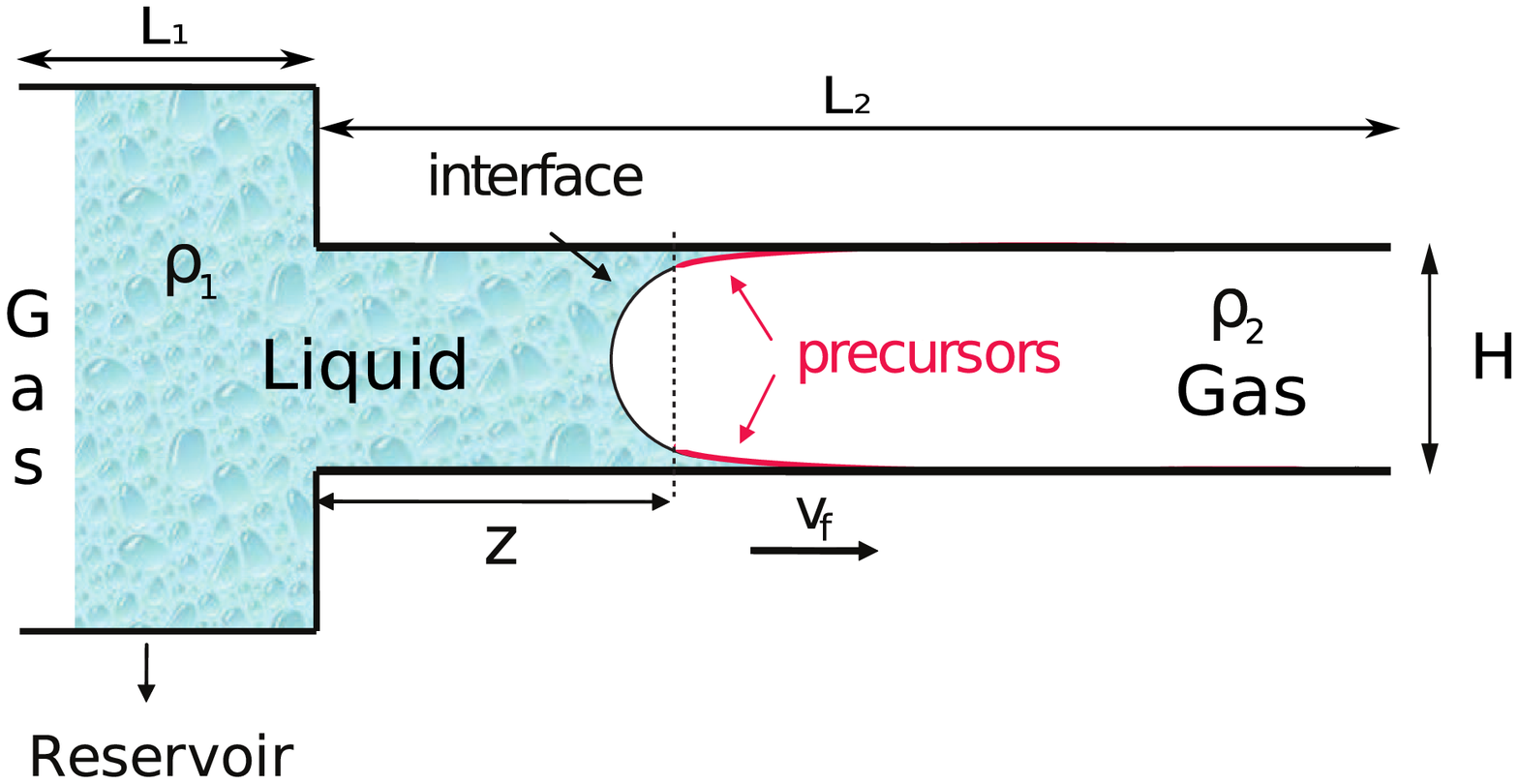} \\ (b)
\includegraphics[scale=0.45]{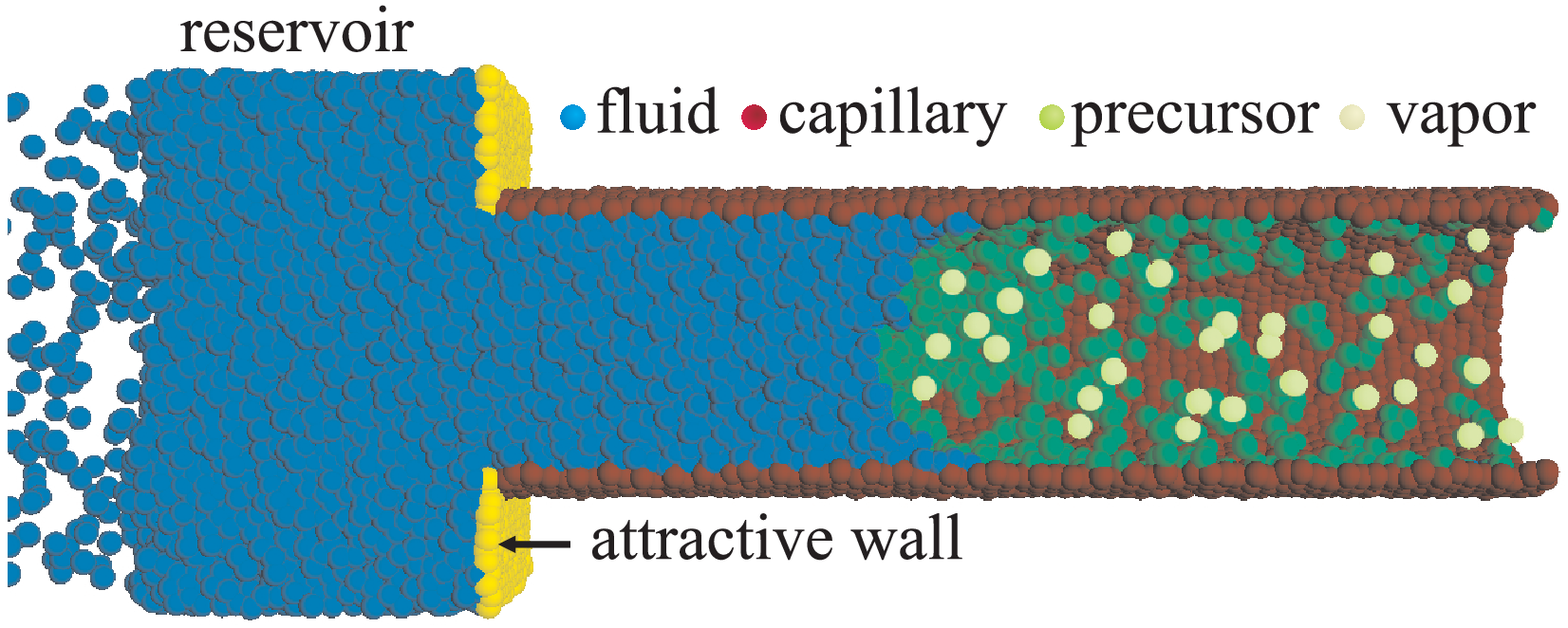}
\end{center}
\caption{Sketch  of  the geometry  used  for  the  description of  the
capillary  imbibition  in  the  LB  and MD  simulations.   (a)  The  2
dimensional geometry, with length  $L_1+L_2$ and width $H$, is divided
in  two parts.  The  left part  has top  and bottom  periodic boundary
conditions, so  as to support  a perfectly flat  gas-liquid interface,
mimicking  an ``infinite  reservoir''. In  the right  half,  of length
$L_2$,  there is  the actual  capillary: the  top and  bottom boundary
conditions   are   solid   wall,    with   a   given   contact   angle
$\theta$. Periodic  boundary conditions are also imposed  at the inlet
and  outlet  sides.   The  main  LB parameters  are:  $H\equiv  ny=40,
L_2=nz=170,    \rho_1=1;    \rho_2=0.35,   \mu_1=0.66,    \mu_2=0.014,
\gamma=0.016$ where  $H$ is the  channel height, $L_2$ is  the channel
length,   $\rho_2$  and   $\rho_1$  the   gas  and   liquid  densities
respectively: $\mu_k$,  $k=1,2$ the dynamic  viscosities, and $\gamma$
the surface  tension. (b) Snapshot of  fluid imbibition for  MD in the
capillary at time $t=1300$ MD  time-steps. The fluid is in equilibrium
with its vapour. Fluid atoms are in blue. Vapour is yellow, tube walls
are red and  the precursor is green. One  distinguishes between vapour
and precursor, subject to the  radial distance of the respective atoms
from the  tube wall,  if a  certain particle has  no contact  with the
wall,  it  is deemed  'vapour'.   The  MD  parameters are  as  follows
\cite{Dim_07}:    $R=11    \sigma,    L=80    \sigma,    \rho_l=0.774,
~\mu=6.3,~\gamma=0.735,~\sigma=1$, where  $R$ is the  capillary radius
and $L$ its length.  }
\label{Fig:1}
\end{figure}

We consider a capillary filling  experiment, whereby a dense fluid, of
density  and  dynamic  viscosity  $\rho_1,\mu_1$,  penetrates  into  a
channel   filled  up   by   a  lighter   fluid,  $\rho_2,\mu_2$,   see
fig. \ref{Fig:1}.  For this kind of fluid flow, the Lucas-Washburn law~\cite{Wash_21,Lucas}
is     expected     to    hold,     at     least    at     macroscopic
scales  and   in  the  limit   $\mu_1\gg  \mu_2$.
Recently,  the   same  law  has  been  observed   even  in  nanoscopic
experiments  \cite{Hub_07}.  In  these limits,  the LW equation  governing the
position, $z(t)$ of the macroscopic  meniscus reads: \be z^2(t) - z^2(0)
= \frac{\gamma  H cos(\theta)}{C \mu }  t , \ee where  $\gamma$ is the
surface tension between  liquid and gas, $\theta$ is  the {\it static}
contact  angle, $\mu$  is the  liquid  viscosity, $H$  is the  channel
height and the factor $C$ depends on the flow geometry (in the present
geometry  $C_{LB}=3\,;\,C_{MD}=2$).   The  geometry  we are  going  to
investigate is depicted in fig.  \ref{Fig:1} for both models.  It is
important to underline that in the LB case, we simulate two immiscible
fluids, without any phase transition.

\begin{figure} \begin{center}\vspace{0.5cm} (a)
\includegraphics[height=5cm,width=7.cm]{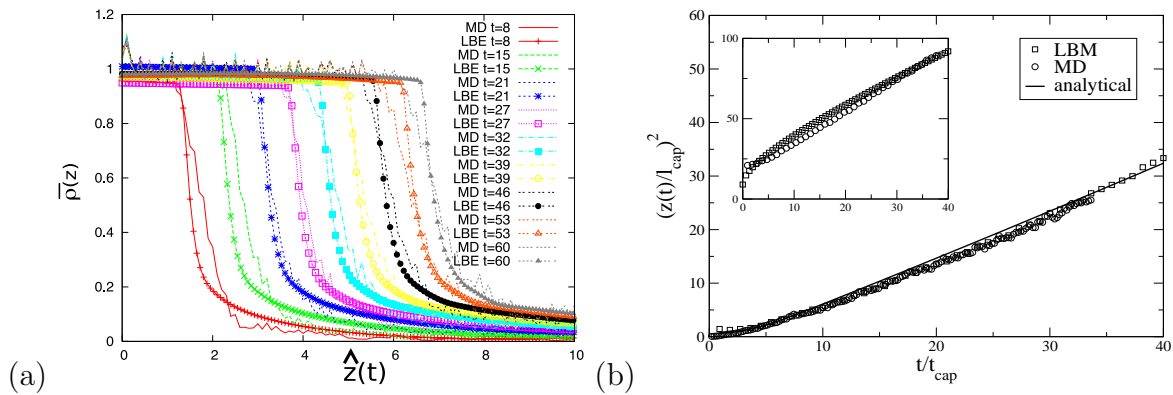}
(b)
\includegraphics[height=5cm,width=7.cm]{fig2b.eps}
\end{center}
\caption{ Dynamics of the bulk and precursor meniscus. (a) Position of
the  liquid meniscus  $\hat{z}^2(t)$ for  LB and  MD  simulations. The
position of the precursor  film, $\hat{z}^2_{prec}(t)$ is also plotted
for  both  models.   $\hat{z}_{prec}$  is  defined  as  the  rightmost
location with density  $\rho=\rho_{bulk}/3$.  All quantities are given
in natural  ``capillary'' units  (see text). The  asymptotic ($t  > 15
t_{cap}$)  rise of  both precursor  and  bulk menisci  follows   a
$t^{1/2}$  law,  with  different  prefactors (see  the  two  straight
lines),    even  though  the  underlying  microscopic  physics  is
different.  Notably, the precursor film  is found to proceed with the
law: $\hat{z}^2_{prec}(t)=1.35 \hat{t}$.   (b) Profiles of the average
fluid density  $\overline{\rho}(z)$ in the capillary  at various times
for LB and MD models.}
\label{Fig:2}
\end{figure}
\begin{figure}
\begin{center}
\vspace{0.5cm}
\includegraphics[scale=0.3]{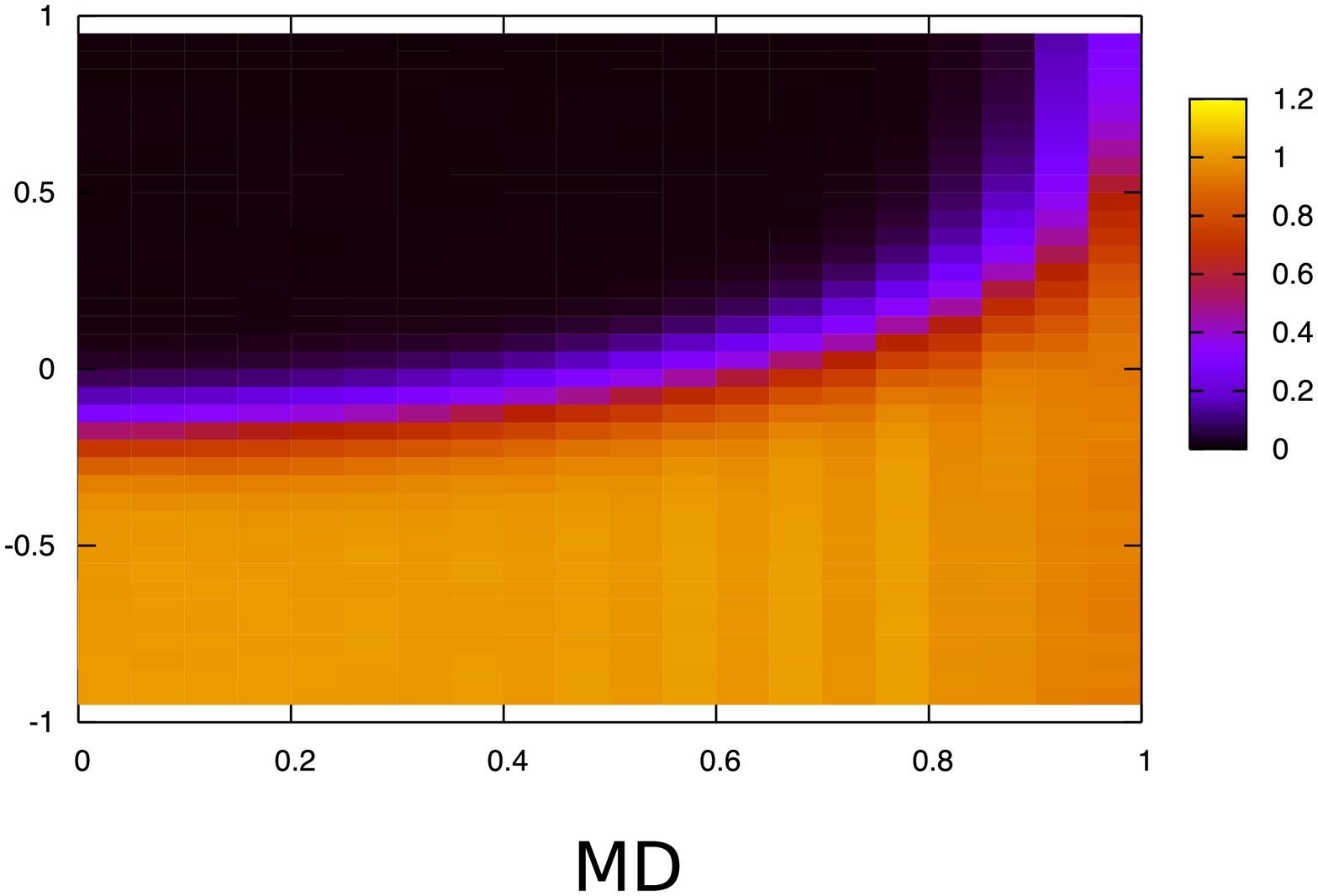}
\includegraphics[scale=0.3]{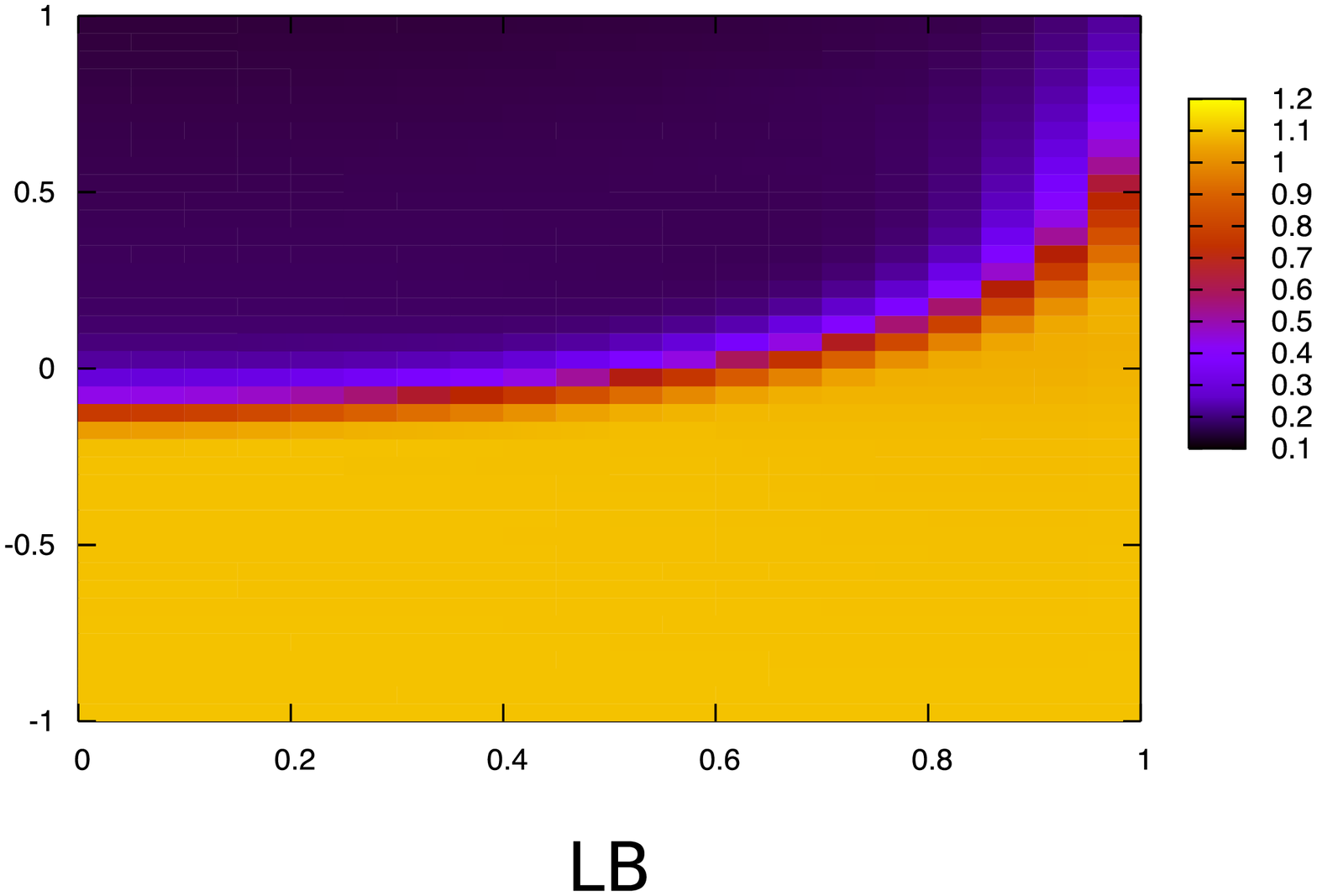}
\caption{The  two  figures  show  the  fluid density  profile  in  the
vicinity of  the meniscus, LJ-MD (left) and LB (right), at
time $\hat{t}=20$ The MD results are rescaled so that the width is the
same for both methods.}
\label{Fig:3}
\end{center}
\end{figure}
Since  binary LB  methods do  not easily  support high  density ratios
between the two  species, we impose the correct  ratio between the two
dynamic viscosities  through an  appropriate choice of  the kinematic
viscosities. The chosen parameters correspond to an average capillary
number  $Ca \approx 3  \; 10^{-2}$  and $Ca  \sim 0.1$  for LB  and MD
respectively.  In  order to emphasise  the universal character  of the
phenomenon and to  match directly MD and  LB profiles, results are
presented in natural units, namely,  space is measured in units of the
capillary  size, $l_{cap}$  and time  in units  of the  capillary time
$t_{cap}=l_{cap}/V_{cap}$, where $V_{cap}=\gamma/\mu$ is the capillary
speed and $l_{cap}=H/C_{LB}$ for LB and $l_{cap}=R/C_{MD}$ for MD. 
The reduced variables are denoted as $\hat{z}$ and $\hat{t}$.

In  figure  \ref{Fig:2}a,  we  show  $\overline{\rho}(z)$  at  various
instants (in capillary units),  for both MD and LB simulations.
Choosing   a  constant   time-interval   $\Delta  \hat{t}=7$   between
subsequent profiles, it is  clear that the interface position advances
slower than  linearly with time.  The relatively  high average density
$\overline{\rho}(z)$  near the  wall witnesses  the
presence  of a  precursor film  attached  to the  wall.  Indeed, the  profiles
$\overline{\rho(z)}$ at late times become distinctly nonzero far ahead
of the interface position (near  the right wall at $\hat{z}\approx 10$
where the  capillary ends), due to  a fluid monolayer  attached to the
wall of the  capillary: this precursor advances faster  then the fluid
meniscus  in the  pore center,  but also  with a  $\sqrt{t}$  law (see
below).  From  this figure, it  is  appreciated  that quantitative
agreement between MD and LB is  found also between the spatial
profiles  of  the density  field.   This  is  plausible, since  the  LB
simulations operate on  similar principles as the MD  ones, namely the
fluid-wall interactions  are {\it not} expressed in  terms of boundary
conditions on the contact angle, like in continuum methods, but rather
in terms of fluid-solid (pseudo)-potentials. In particular, the degree
of hydrophob/philicity of the  two species can be tuned independently,
by choosing different values  of the fluid-solid interaction strengths
(for  details see \cite{Dim_07,Chi_08}).   In figure  \ref{Fig:2}b, we
show the position  of the advancing front and the  precursor film as a
function of  time, for  both MD and  LB simulations.  Even  though the
average capillary  numbers are  not exactly the  same, a  pretty good
agreement  between LB  and MD  is  observed.  In  particular, in  both
cases, the  precursor is   found  to obey a  $\sqrt t$  scaling in
time, although with  a larger prefactor than the  front.  As a result,
the relative distance between the  two keeps growing in time, with the
precursors  serving  as  a  sort  of  'carpet',  hiding  the  chemical
structure of the  wall to the advancing front. In spite of the different capillary number, the agreement between
LB and MD  can be attributed
to  the fact that  these two  approaches share  the same  {\it static}
angle,  $\theta_0=0$,  since  the ``maximal  film''~\cite{DeG_84},  or
complete  wetting, configuration  has been  imposed by  increasing the
strength  of   the  wall-fluid  attraction~\cite{Dim_07,Chi_08},  i.e.
imposing that the spreading coefficient $S>0$~\cite{DeG_84}.

In natural units, the Lucas-Washburn law takes a very simple universal
form
\begin{equation}
\label{eq:univ} {\hat z}^2=\hat{z}_0^2+ \hat{t}~,
\end{equation} where  we have  inserted the value  of $cos(\theta)=1$,
corresponding  to complete wetting.   As to  the bulk  front position,
fig.\ref{Fig:2}a shows that both MD  and LB results superpose with the
law (\ref{eq:univ}),  while the  precursor position develops  a faster
dynamics, fitted by the relation~\cite{Chi_08b}:
\begin{equation} {\hat z}^2_{prec}=\hat{z}_0^2+ 1.35 \hat{t}~.
\label{eq:prec}
\end{equation}  Similar speed-up  of the  precursor has  been reported
also    in   different    experimental   and    numerical   situations
\cite{Bico,Popescu}.   The  precursor  is  here  defined  through  the
density  profile, $\overline{\rho}(z)$,  averaged  over the  direction
across the channel.

In figure \ref{Fig:3}, we show a visual representation of
the quantitative agreement between MD  and LB dynamics: 
we  present  the density  isocontour  which  can  be imagined  as  the
advancing front. It  is seen that the interface  positions are in good
agreement and also density structures are very similar.

These results  achieve the validation of the
LB method  against the MD simulations.  Moreover,  as recently pointed
out~\cite{Dim_07,Hub_07,Chi_08b},    our   findings    indicate   that
hydrodynamics persists down to nanoscopic scales. The fact that the MD
precursor  dynamics  {\it   quantitatively}  matches  with  mesoscopic
simulations, suggests  that the precursor physics also  shows the same
kind of nanoscopic persistence.

\begin{figure}
\begin{center}\vspace{1.cm}
\includegraphics[scale=0.4]{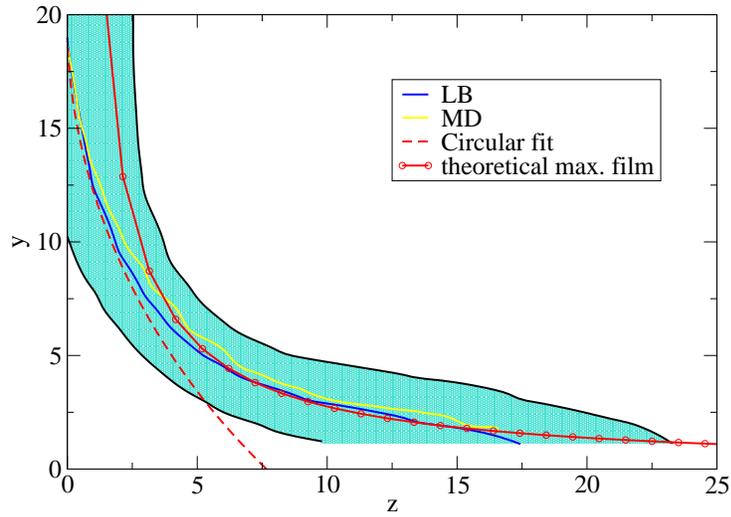}\vspace{0.1cm}
\caption{The extrapolated  interface position for  both methods and
the theoretical profile  for ``maximal film''. The x-  and y- axis are
normalised in  order to reproduce  the LB units.  The  colored region
represents the whole  interface. It is delimited, on  the left, by the
solid line representing the  isolines of the bulk density ($\rho=0.9$)
and on  the right by the  one describing the beginning  of the lighter
fluid ($\rho=0.1$), where only  few molecules of the penetrating fluid
are present  because of  diffusion. The length  of the  precursor film
appear in line  with the relation $L_p=7.2 10^{-1}Ca^{-1}$  found in a
recent nanoscale experiment~\cite{Kav_03}.}
\label{Fig:4a}
\end{center}\end{figure}
\begin{figure}
\begin{center}\vspace{0.5cm}
\includegraphics[scale=0.8,angle=-90]{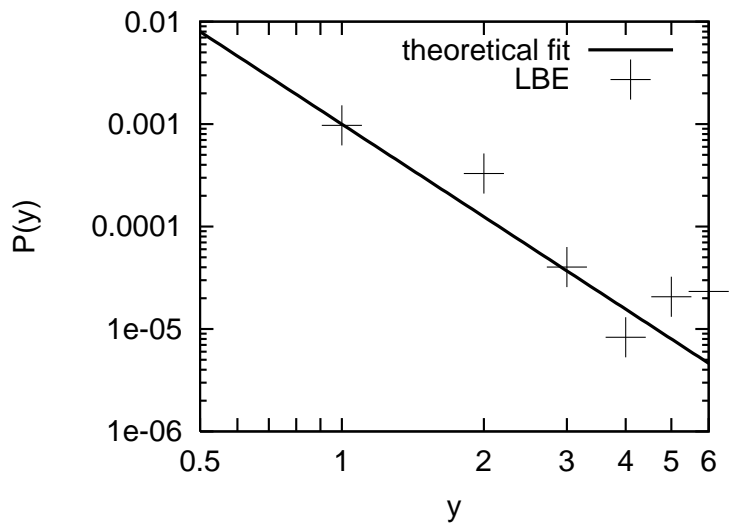}
\caption{ Disjoining pressure in
LB and its fit based  upon analytical expression are shown versus the
distance from wall. The analytical expression for Van der Waals forces
is given by $\frac{A}{6\pi y^3}$, where is the Hamakar constant $A=\pi
^2k\alpha_l(\alpha_s-\alpha_L)$   and   $\alpha_s,\alpha_L$  are   the
polarizabilities of  the solid  and the liquid.  In our case  of total
wetting,  $A>0$~\cite{DeG_book}  and  our  data are  fitted  by  using
$A=0.019$.}
\label{Fig:4b}
\end{center}\end{figure}   Fig.~\ref{Fig:4a} shows the  whole interface
in the vicinity  of the meniscus and, in particular,  the shape of the
precursor computed with both methods.  The plot emphasises the presence
of a precursor film at the wall. The agreement between the two methods
is again quantitative.  The interface in both methods  turns out to be
about $5$ units.   The precursor film profile, defined  by the isoline
of  points   with  density  $\rho=\rho_{bulk}/3$,  is   split  in  two
regions. In the first that  arrives until $y\approx7$, it is fitted by
a circular profile with center at $z_0=0,y_0=20$ and radius $r=18$. In
the second,  from $y\approx7$ to the  wall, the film is  fitted by the
function $\frac{a^2}{Ca z}$. In  this formula, $a$ is a characteristic
molecular size which  is taken to be $a=1$ in  our case.  This profile
is obtained in the lubrication approximation considering Van der Waals
interaction  between fluid  and walls  \cite{DeG_84}, in  the  case of
``maximal film''  (perfect wetting).  Quantitative  agreement is again
found between LB, MD and analytics, thus corroborating the idea of the
Shan-Chen pseudo-potential as a {\it quantitative} proxy of attractive
Van der Waals interactions in  the low density regime, where hard-core
repulsion  can be  neglected.   It  is interesting  to  note that  the
presence  of   the  precursor   film  guarantees  an   apparent  angle
$\theta=0^{\circ}$,whereas the angle  calculated from the circular fit
of the  bulk meniscus  would be $\theta\approx40^{\circ}$  Since these
forces are, apparently, the  key microscopic ingredient to be injected
into an otherwise continuum framework,  it is plausible to expect that
LB  should  be  capable  of  providing a  realistic  and  quantitative
description  of  precursor  dynamics.   The  results  of  the  present
simulations  confirm   these  expectations   and  turn  them   into  
quantitative evidence.  We have  also computed the disjoining pressure
inside the film in order to corroborate the idea that LB is capable of
correctly  describing  the  dynamics  of  the  precursor  film.   Such
pressure is  related to the chemical  potential and it  is function of
the  distance   from  walls,   that  is  of   the  thickness   of  the
film~\cite{DeG_book}.  In fig.   \ref{Fig:4b}, the disjoining pressure
computed inside the film in the LB is compared with the fit based upon
the  analytical expression given  for Van-der-Waals  forces.  LB  is a
diffuse-interface  model  and, therefore,  can  not  guarantee a  pure
hydrodynamical  behaviour with  ideal interfaces  inside a  thin film,
such as the precursor film  experienced in this work. Nevertheless, is
seems  to  assure  a  disjoining  pressure at  least  compatible  with
microscopic physics,  with the correct  divergence at the  wall.  This
appears to be consistent with  the fact that real interfaces are found
to be diffuse also in experiments at macroscopic scales~\cite{Bico}.

\subsection{Nanochannel in presence of  an obstacle}

\begin{figure} \begin{center} \vspace{0.5cm} 
\includegraphics[scale=0.3]{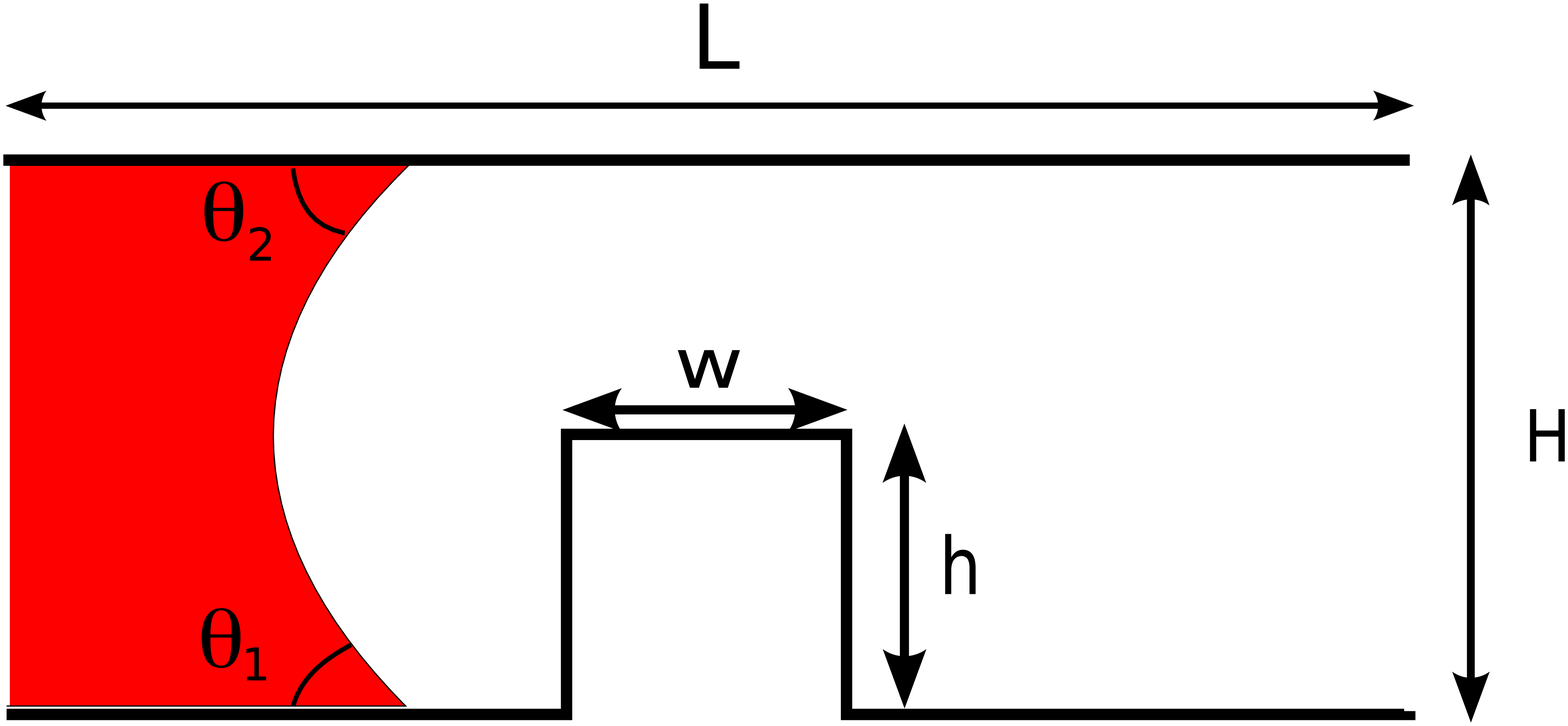}
\caption{A  channel of  length $L=200$  LB units  and width  $H=41$ LB
units is studied.  The contact angles at the bottom  and top walls are
taken  to be  $\theta_1=\theta_2=\theta$. An  square obstacle  is also
present, whose  dimensions are given by  the height $h$  and the width
$w$.}
\label{Fig:5}
\end{center}\end{figure}
\begin{figure} 
\begin{center} \vspace{0.5cm} \hspace{-2.cm}
\includegraphics[height=9cm,width=14.cm]{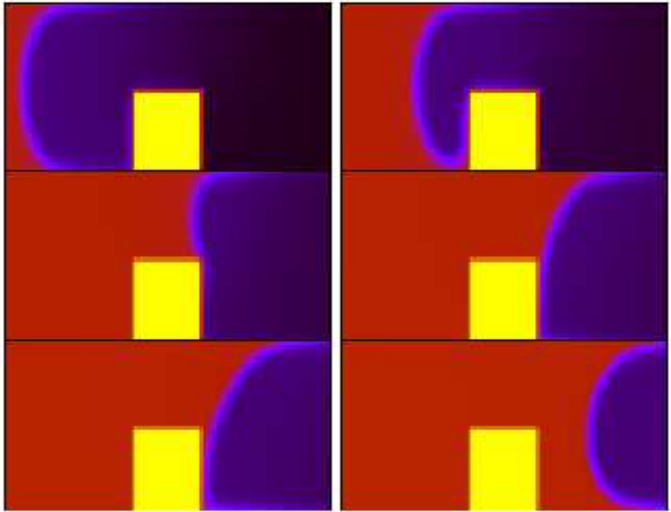}
\end{center}
\caption{Snapshots      of     density     at      different     times
$t=20000,30000,40000,50000,60000,100000$ LB units.}
\label{Fig:6}
\end{figure}
\begin{figure} \begin{center} \vspace{0.5cm} (a)
\includegraphics[height=6cm,width=9.cm]{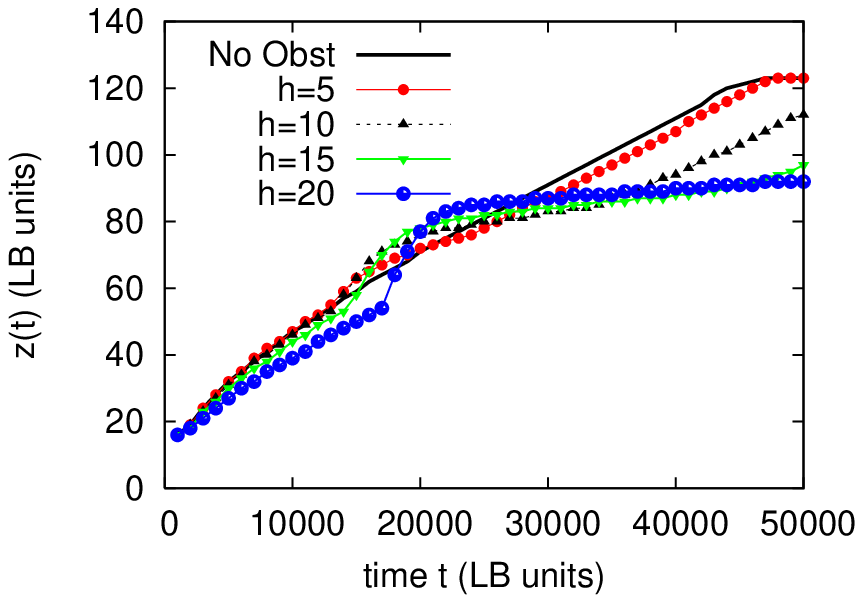} \\ (b)
\includegraphics[height=6cm,width=9.cm]{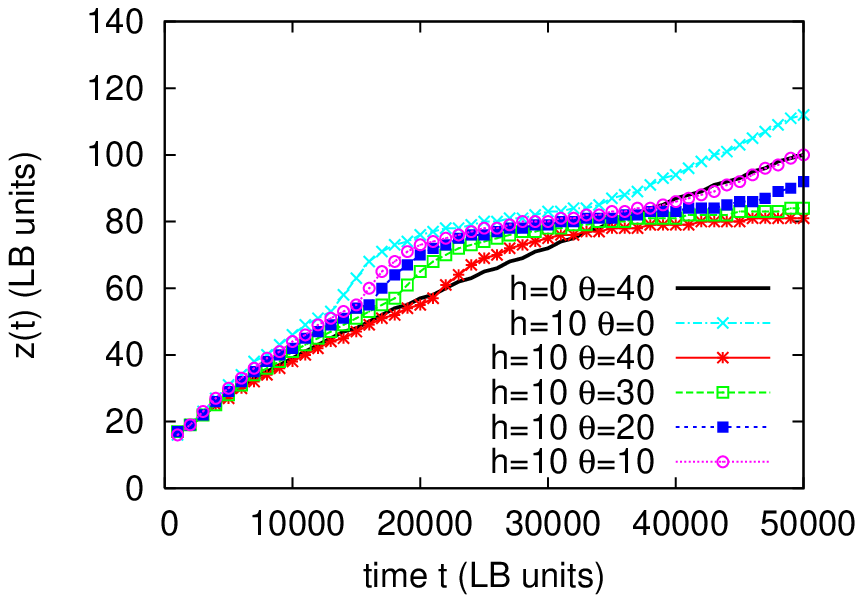} \\ (c)
\includegraphics[height=6cm,width=9.cm]{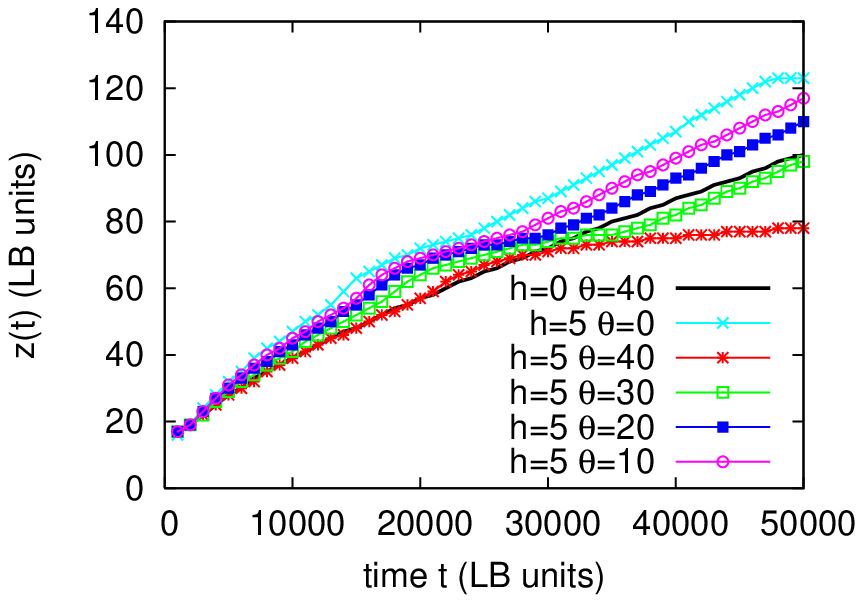}
\end{center} \caption{In  all figures  square obstacles  (with width  equal  to the
height) are  considered and the  front position versus time  is shown.
(a)     Different    obstacle     height     are    considered     for
$\theta=0^{\circ}$.  (b)The contact  angle is  changed (from  0  to 40
degrees)  taking the  height of  the  obstacle constant  and equal  to
$h=10$.  The  front  dynamics  for  the  case  with  no  obstacle  and
$\theta=40^{\circ}$ is also  shown for comparison. (c) Same  as in (b)
but with $h=5$.}
\label{Fig:7}
\end{figure}  

In this section,  we study the capillary filling  in a nano-channel in
presence of an obstacle. In  particular, we want to analyse the effect
of the precursor films on the  dynamics of the fluid when crossing the
obstacle.  This test-case has been  simulated through the LB model and
the geometry is depicted in  figure \ref{Fig:5}.  First, we simulate a
channel  with  hydrophilic  walls  which support  a  complete  wetting
($\theta=0^{\circ}$). An obstacle is put  in the middle of the channel
length with dimensions $w=h=20$ LB units. Six snapshots of the density
evolution with time are  represented in figure \ref{Fig:6}.  Precursor
films are  visible ahead of the  meniscus of the  penetrating fluid at
the bottom wall. Such film  wets the obstacle before the front encounters it.
 The obstacle  is large (its  width is equal  to the
half of  the channel) but the  contact angle is small  and therefore the
front does  not pin  and is able  to pass  the obstacle in  a relative
short time, as known according to the Gibbs, or Concus-Finn, criterion~\cite{Con,Gibbs}.
Nevertheless, the  meniscus dynamics is deeply affected by
the presence of the obstacle. It is seen that the liquid on the
top  wall  advances much  more  rapidly than  the  one  on the  bottom
wall. In particular,  at the beginning and at the  end of the obstacle
the meniscus line is strongly distorted. 
After some time, 

We now perform  a systematic series of simulations,  where we vary the
height of the  obstacle in order to check to  what extent the dynamics
is affected  by wall roughness.   In figure \ref{Fig:7},  the detailed
characteristics  of  the  front  dynamics  are  analysed.   In  figure
\ref{Fig:7}a, it is possible  to appreciate that, for complete wetting
$\theta=0^{\circ}$, the  presence of  a small obstacle  ($h=5=H/8$) is
negligible.  The front  advances almost in the same  way for the cases
$h=0^{\circ}$ and $h=5$. This is possibly due to the presence in these
cases  of precursor  films  which wet  the  obstacle anticipating  the
meniscus. In this  way, they ``hide'' the roughness  and let the front
pass  through the  obstacle  without feeling  any  roughness.  For  an
obstacle of middle  width $h=10$, the front is  strongly distorted and
slowed down  but it  recovers the initial  slope, after it  has passed
through the obstacle. On the  contrary, for larger obstacles $h\ge 15$
the dynamics of  the meniscus  seems to be irreversibly changed  and it advances
with  a  decreased slope  after  the  obstacle. 
Then, we  carry  out
numerical  simulations keeping  the  height of  the obstacle  ($h=10$)
constant, while  varying the  contact angle.  In  figure \ref{Fig:7}b,
it  is seen
that the slope  of the curve is strongly affected by  the value of the
contact angle  and, notably, for $\theta>20^{\circ}$ the  curve is not
able  to recover  the initial  dynamics  but it  results much  slower.
Moreover, the  velocity of the  advancing front is  strongly decreased
when  the walls  are not  very  hydrophilic and  with roughness.   For
instance,  the front advances  very similarly  for $\theta=40^{\circ}$
without roughness $h=0$  and for $h=10,\theta=10^{\circ}$.  This seems
to confirm the  guess that the precursor films can  reduce the drag in
presence of  roughness.  In figure \ref{Fig:7}c, the  same analysis is
worked  out for  an obstacle  with width  $w=5$.  The  effect  of such
obstacle is quite small for $\theta<10^{\circ}$, where precursor films
are  supposed  to  be  present.  Naturally,  it  results  increasingly
important  with the  increase of  the contact  angle and  it  causes a
change in slope for $\theta>30^{\circ}$.

\section{Conclusions}

Summarising, we  have thoroughly analysed  the capillary filling  in a
nanochannel by using atomistic and hydrokinetic methods.

We  have  reported quantitative  evidence  of  the  formation and  the
dynamics of precursor films  in capillary filling with highly wettable
boundaries.  The  precursor shape  shows persistent deviation  from an
ideal circular  meniscus, due to the nanoscopic  distortion induced by
the  interactions with  the walls.  
 When properly scaled, the results do not seem sensitive to geometry  (in this work we investigate two different geometries) and resolution. 
 This  has been  connected to  the
disjoining  pressure induced  by  Van der  Waals interactions  between
fluid  and solid and  approximated in  the LB  approach by  a suitable
phenomenological  model.    Our  findings  are   supported  by  direct
comparison  between LB,  MD simulations  and  theoretical predictions,
which  suggests that  a  continuum description  (LB)  together with  a
proper inclusion  of the solid-fluid interaction is  able to reproduce
this  phenomenon.   In  this  sense,  this work  provides  a  complete
assessment of  the LB  method for the  study of  nanochannel capillary
filling. 

Then, a nanochannel with a square obstacle has been  simulated via the LB model. The complete
parametric  analysis  points  out  that  for   highly  wetting  walls
($\theta<10^{\circ}$)  the presence of  precursor films  make the  presence of
small obstacles (with a width  smaller than 1/8 of the channel height)
negligible. Furthermore,  results seems to indicate  that to reach
high flow rate  it is preferable to choose  very hydrophilic and rough
walls rather than to use very smooth but hydrophobic walls.

3-D  simulations  will  be  carried  out  in  order  to  assess  these
affirmations and to evaluate the effect of different obstacle geometry
and position.

\section{acknowledgments} S. Chibbaro's work  is supported by a ERG EU
grant.  This work  makes use of results produced  by the PI2S2 Project
managed by  the Consorzio COMETA,  a project co-funded by  the Italian
Ministry of  University and Research (MIUR).   He greatly acknowledges
the  financial  support given  also  by  the  consortium SCIRE.   More
information  is   available  at  http://www.consorzio-cometa.it.  Work
performed under the NMP-031980 EC project (INFLUS).

\end{document}